\documentclass[...]{aa501}

\usepackage{psfig}

\newcommand{\Hei}{He~{\sc i}}
\newcommand{\Heii}{He~{\sc ii}}
\newcommand{\hei}{He~{\sc i} $\lambda$4471}
\newcommand{\heii}{He~{\sc ii}  $\lambda$4542 }

\newcommand{\teff}{\ifmmode T_{\rm eff} \else $T_{\mathrm{eff}}$\fi}
\newcommand{\logg}{\ifmmode \log g \else $\log g$\fi}
\newcommand{\msun}{\ifmmode M_{\odot} \else M$_{\odot}$\fi}
\newcommand{\zsun}{\ifmmode Z_{\odot} \else Z$_{\odot}$\fi}
\newcommand{\lsun}{\ifmmode L_{\odot} \else L$_{\odot}$\fi}
\def\aap{A\&A}
\def\aaps{A\&AS}
\def\aas{A\&AS}

\def\apj{ApJ}

\def\mnras{MNRAS}

\begin{document}



\title{On the effective temperature scale of O stars}

\author{Fabrice Martins 
           \inst{1} 
        \and Daniel Schaerer 
           \inst{1}
        \and D.\ John Hillier
           \inst{2}}

\offprints{F. Martins, martins@ast.obs-mip.fr \inst{1}}

 \institute{Observatoire Midi-Pyr\'en\'ees, Laboratoire d'Astrophysique, 
 14, Av.  E. Belin, F-31400 Toulouse, France 
 \and
 Department of Physics and Astronomy, University of Pittsburgh,
 Pittsburgh, PA 15260, USA}

\date{A\&A, in press}

\titlerunning{Temperature scale of O stars}

\abstract{We rediscuss the temperature of O dwarfs based on new non-LTE 
line blanketed atmosphere models including stellar winds computed 
with the {\em CMFGEN} code of Hillier \& Miller (\cite{hm98}).
Compared to the latest calibration of Vacca et al.\ (\cite{vacca}),
the inclusion of line blanketing leads to lower effective temperatures,
typically by $\sim$ 4000 to 1500 K for O3 to O9.5 dwarf stars.
The dependence of the \teff--scale on stellar and model parameters
-- such as mass loss, microturbulence, and metallicity -- is explored,
and model predictions are compared to optical observations of O stars.
Even for an SMC metallicity we find a non-negligible effect of line
blanketing on the \teff--scale.
The temperature reduction implies downward revisions of luminosities
by $\sim$ 0.1 dex and Lyman continuum fluxes $Q_0$ by approximately
40\% for dwarfs of a given spectral type.
\keywords{Stars: general  -- Stars: temperature -- Stars: fundamental parameters 
-- Stars: atmospheres}}

\maketitle

\section{Introduction}
\label{s_intro}

As a significant fraction of the flux of O stars is emitted in
the inaccessible Lyman continuum ($\lambda < 912$ \AA) reliable direct
determinations of their effective temperatures are not possible.
Indirect methods, primarily based on atmospheric modeling, are
therefore employed (e.g.\ B\"ohm-Vitense \cite{bv81}, Crowther \cite{pac98}).
Given the need for a detailed treatment of non-LTE effects and the presence 
of stellar winds (Kudritzki \& Hummer \cite{kh90}), 
a complete modeling of such atmospheres including also the effects of
numerous metal-lines (``line blanketing'') remains a complex task
(cf.\ Schaerer \& Schmutz \cite{ss94}, Hillier \& Miller \cite{hm98}, 
Pauldrach et al.\ \cite{p01}).

For these reasons, most published spectral analysis have so far been based 
on simple non-LTE models.
For example, the most recent calibration of stellar parameters of O and early 
B type stars of Vacca et al.\ (\cite{vacca}, hereafter VGS96) 
is based only on results from
plane parallel, pure Hydrogen and Helium (H-He) non-LTE models.
Their derived temperature scale for O stars is found to be significantly
hotter than most earlier calibrations (see references in VGS96).
Such differences lead to non-negligible changes in the fundamental
parameters of O stars --- e.g.\ luminosities, Lyman continuum fluxes etc.\
--- when estimated from spectral types.
Accurate calibrations are crucial for various astrophysical topics, such
as comparisons with stellar evolution models, determinations of the initial 
mass function and cluster ages, studies of H~{\sc ii} regions, and others.

Indications for a decrease of \teff\ due to line blanketing effects
have been found since the first non-LTE $+$ wind modeling attempts
by Abbott \& Hummer (\cite{ah85}, and subsequent investigations
based on the same ``wind blanketed'' models), the improved models 
of Schaerer \& Schmutz (\cite{ss94}) and Schmutz (\cite{s98}),
and the fully-blanketed plane parallel non-LTE models of
Hubeny et al.\ (\cite{hub98}).
Similar indications are obtained by Fullerton et al.\ (\cite{ful00})
from recent modeling of {\em FUSE}\, spectra with the code of Pauldrach 
et al.\ (\cite{p01}) and by Crowther et al.\ (\cite{cro01}).

The effective temperature scale of O stars is revised here
based on the recent {\em CMFGEN} code of Hillier \& 
Miller (\cite{hm98}), which treats the problem of a non-LTE line 
blanketed atmosphere with a stellar wind in a direct way,
thereby avoiding possible shortcomings due to opacity sampling
techniques employed by Schaerer \& Schmutz (\cite{ss94}), 
Schmutz (\cite{s98}), and Pauldrach et al.\ (\cite{p01}).
First results on the dwarf sequence are presented here.
A more detailed account including all luminosity classes will
be presented in a subsequent publication.

In Sect.\ 2 we describe our method and the calculated models.
The results, their dependence on model/stellar parameters, and
first comparison with observations are presented in Sect.\ 3. 
Implications of the revised \teff\ scale and remaining uncertainties
are discussed in Sect.\ 4.

\section{Model ingredients}
\label{s_models}

We have constructed spherically expanding non-LTE line-blanketed model atmospheres 
using the {\em CMFGEN} comoving-frame code of Hillier \& Miller (\cite{hm98}).
This code solves the equations of statistical equilibrium, radiative transfer, and 
radiative equilibrium, and allows for a direct treatment of line blanketing
through the use of a super-level approach.
The following ions are included in our calculations:
H, He~{\sc i-ii}, C~{\sc ii-iv}, N~{\sc ii-v}, O~{\sc ii-vi}, Si~{\sc ii-iv},
S~{\sc iv-vi}, and Fe~{\sc iii-vii}, whose $\sim$ 2000 levels are described by 
$\sim$ 700 super-levels, 
corresponding to a total of $\sim$ 20000 bound-bound transitions.

For simplicity a constant Doppler profile
(thermal width corresponding to the mass of Helium and $T=20000$ K
plus a microturbulent velocity of $v_{\rm turb}=20$ km/s)
is assumed for all lines in the statistical 
equilibrium and radiative transfer computation.
To examine if a constant thermal width and
the use of the large microturbulent velocity does not artificially enhance
the photospheric blanketing, we have made 
test calculations with the correct depth and ion dependent thermal width
and $v_{\rm turb}=0.1$ km/s.
No significant changes in atmospheric structure, level populations, and the 
emergent spectrum were found.
This is explained in part by the high density of lines in the UV part of the spectrum,
which implies an average spacing between lines which is smaller than the 
typical Doppler width. The opacity in the wing of a line is therefore 
mostly dominated by the core opacity of the neighbouring line, and the exact 
intrinsic line profile is of little importance.
With our standard choice, $\sim$ 80000 frequency points are necessary to 
correctly sample all lines.

The input atmospheric structure, connecting smoothly the spherically extended hydrostatic layers with
the wind (parametrised by the usual $\beta$-law), is calculated as in Schaerer \& de Koter (\cite{sdk})
with the {\em ISA-WIND} code of de Koter et al.\ (\cite{dk96}) 
As the approximate temperature structure in {\em ISA-WIND} differs from the final 
radiative equilibrium temperature structure, the atmosphere structure in the quasi-hydrostatic 
part may be inconsistent with the final gas pressure gradient.
However, for the issues discussed here the differences are small (corresponding to a change of 
$\la$ 0.1 dex in $\log g$). In any case, the lines considered here are formed
in the transition region whose structure/dynamics remain largely parametrised. 
The formal solution of the radiative transfer equation yielding the detailed emergent spectrum 
allows for incoherent electron scattering and includes standard Stark broadening 
tables for H, \Hei, and \Heii\ lines. 
Our standard calculations assume $v_{\rm turb}=5$ km/s.

We have computed a grid of models representative of O dwarfs in the temperature 
range between $\sim$ 30000 and 50000 K. The model parameters are taken from the
{\em CoStar} models A2-E2 of Schaerer \& de Koter (\cite{sdk}), with an additional 
model 
Y2 at (\teff,$\log g$) $\sim$ (31500,4.0) and the remaining parameters 
\footnote{$M=16.83$ M$_{\odot}$, $\log{\teff}=4.498$, $\log(L/L_{\odot})=4.552$, 
$R=6.358$ R$_{\odot}$, $\log{\dot{M}}=-7.204$ M$_{\odot}$/yr, and $v_\infty=2500$ km/s.}
taken from stellar tracks of Meynet et al.\ (\cite{meynet94}).
For each parameter set a line blanketed model with solar metallicity and a pure H-He model
was computed.

\section{Results}

\subsection{Blanketing effect on the temperature scale}
\begin{figure}
\centerline{\psfig{file=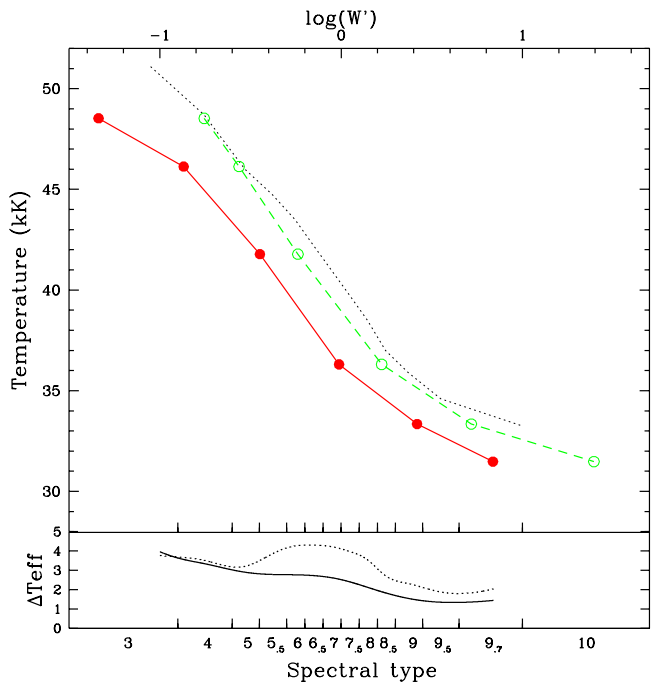,width=9.2cm}}
\caption{{\em Upper panel:} 
Effective temperature of O dwarfs as a function of the spectral subtype (lower scale).
The correspondance between $\log W^\prime$ (upper scale) and spectral type is 
given by Mathys (1988).
For values $\log W^\prime >$ 1.0 we assign a spectral type of 10
Filled circles show our line blanketed models, open circles pure H-He models.
The VGS96 relation (dotted line) is well reproduced by our pure H-He models.
{\em Lower panel:} 
\teff\ shift between H-He and line blanketed models (solid line) and
between VGS96 scale and our line blanketed models (dotted line).
Note the decrease of \teff\ due to line blanketing.
}
\label{fig_scale}
\end{figure}

The optical \hei\ and \heii\ classification lines are used to assign 
spectral types to our models.
Fig.\ \ref{fig_scale} shows the effective temperature as a function of 
$\log W^\prime \equiv \log W(4471) - \log W(4542)$ and the corresponding
spectral type according to Mathys (\cite{mat88}).
The pure H-He models (open circles) follow closely the \teff--scale for dwarfs of 
VGS96, which is based on a compilation of
stellar parameters determined using pure H-He plane parallel non-LTE 
model atmospheres. The comparison shows that if we neglect line blanketing
our dwarf model grid would yield nearly the same absolute \teff--scale as 
the pure H-He plane parallel models adopted for the spectral analysis
included in the compilation of VGS96.

The line blanketed model sequence (Fig.\ \ref{fig_scale}, filled symbols)
shows a systematic shift to earlier spectral types for a given temperature,
or equivalently a shift to {\em lower \teff\ for line blanketed models
at a given spectral type.}
The difference ranges from $\sim$ 1500 K at spectral type O9.5 to $\sim$4000 
K at spectral type O3 (cf.\ Fig.\ \ref{fig_scale}, solid line in lower panel). 
The difference with the VGS96 scale is shown as the dotted line.
Our line blanketed scale smoothly joins earlier calibrations at 
O9.7V (see VGS96, Fig.\ 1).

\begin{figure}[t]
\centerline{\psfig{file=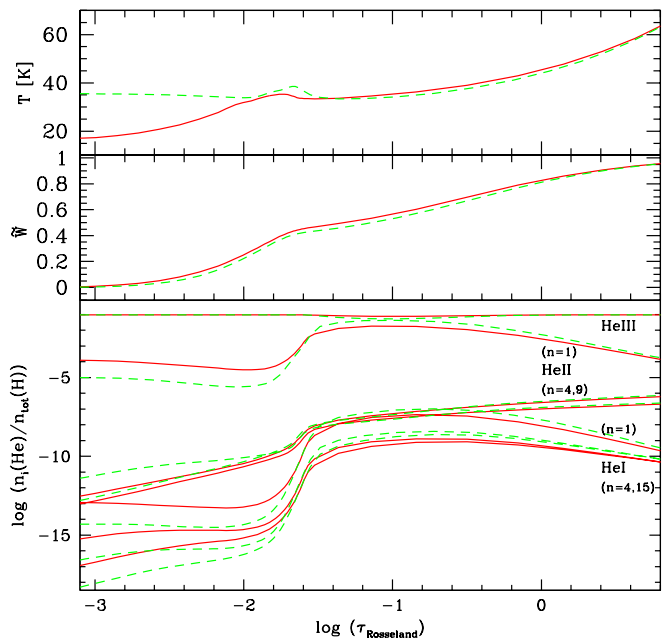,width=9.2cm}}
\caption{Comparison of atmosphere structures of model C2 (\teff=41.8 kK, $\log g=$4.0).
Solid line is for the line-blanketed model and dashed line for the pure H-He model.
{\em Upper panel:}
Temperature structure.
{\em Middle panel:}
Dilution factor $\tilde{W} = 1 - \frac{1}{4} F/J$ where F is the flux 
and J the frequency averaged mean intensity (cf.\ Schaerer \& Schmutz \cite{ss94}). 
{\em Lower panel:}
Populations of the ground levels of Helium and of the lower and upper levels of the 
transitions \hei\ and \heii. Given are the relative number population n$_i$ with 
respect to total H population n$_{tot}$(H). 
}
\label{fig_lb}
\end{figure}

As a spectral type corresponds to a given ionisation state of Helium in the line
formation region, blanketed models must be more ionised than unblanketed models. 
The introduction of line blanketing leads to three main effects 
illustrated in Fig.\ \ref{fig_lb} for the case of model C2 (cf.\ Figs.\ 13 and
14 of Schaerer \& Schmutz \cite{ss94}).
Qualitatively the same trends are obtained for all models.

\begin{figure}
\centerline{\psfig{file=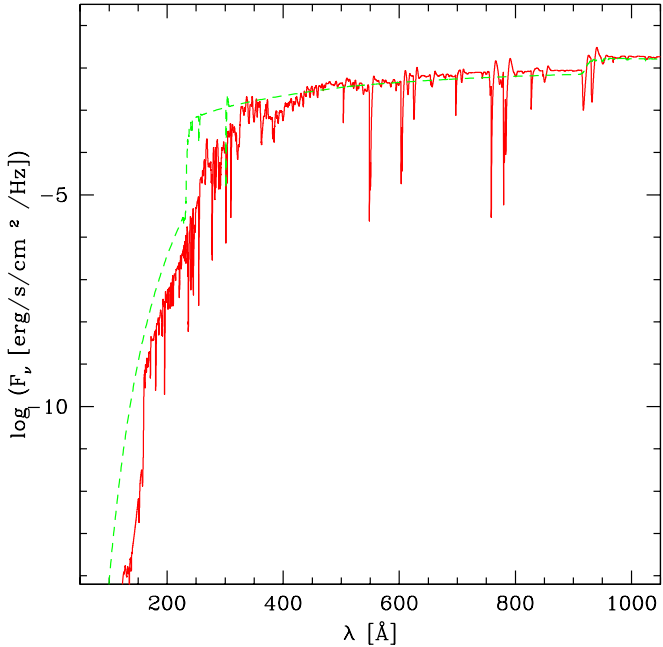,width=9.2cm}}
\caption{UV spectrum of model C2 with line blanketing (solid line) and pure H-He model 
(dashed line). Note the reduction of the EUV flux below $\sim$ 500 \AA\ due to the inclusion 
of metals.
}
\label{fig_sed}
\end{figure}

\begin{itemize}
\item[{\em 1)}] Blanketing leads to the backscattering of photons towards the inner atmosphere 
which forces the local temperature to rise so that flux conservation is fulfilled 
(backwarming effect; see upper panel).
\item[{\em 2)}] At the same time the radiation field becomes more diffuse, as quantified 
by the dilution factor $\tilde{W} = 1 - \frac{1}{4} F/J$ shown in the middle panel, 
causing an increase of the mean intensity
(cf.\ Abbott \& Hummer \cite{ah85}, Schaerer \& Schmutz \cite{ss94}).
\item[{\em 3)}] In the outer part of the atmosphere 
($\log \tau_{\rm Ross} \la$ --2  in Fig.\ \ref{fig_lb}) 
the ionisation is essentially controlled by the EUV flux, which is quite strongly 
reduced due to the blocking by numerous metal lines shown in Fig.\ \ref{fig_sed}.
Here this effect dominates over 2), in contrast with the finding of Schaerer \& Schmutz 
(\cite{ss94}), leading to a lower ionisation. 
\end{itemize}

Effects 1) and 2) lead to a higher ionisation in the formation region of the 
classification lines. 
This results predominantly in an increase of W(4542) at \teff\ $\la$ 38000 K and a 
decrease of W(4471) at higher \teff (cf.\ Fig.\ \ref{fig_obs}).

Given the stronger mass loss and the corresponding increase of the wind density, 
one expects even larger temperature differences between non-blanketed and 
line blanketed models for giant and supergiant luminosity classes (cf.\ Abbott 
\& Hummer \cite{ah85}, Schmutz \cite{s98}, Crowther et al.\ \cite{cro01}).

\subsection{Dependence on model and stellar parameters}
\label{s_depend}
How strongly do our results depend on poorly known parameters such as the 
velocity law in the photosphere-wind transition zone, $v_{\rm turb}$,
and variations of gravity and $\dot{M}$ expected within the dwarf class ?
Do our calculations still miss opacity sources ?

As pointed out by Schaerer \& Schmutz (\cite{ss94}) changes in He line profiles 
due to modifications of the velocity law $v(r)$ in the photosphere--wind transition zone
can lead to similar equivalent widths variations as line blanketing. 
Test calculations for models A2 and C2 varying the slope $\beta$ from 0.8 (our standard value) 
to 1.5 show that both H-He and line blanketed models exhibit a similar shift
in $\log (W^\prime)$. The obtained {\em relative} \teff\ difference
between H-He and blanketed models remains thus identical.
The blanketed models with $\beta=1.5$ have $\log(W^\prime)$ lowered by $\sim$ 0.1--0.2 dex.
However, as H$\alpha$ profile fits for O dwarfs are generally quite compatible with 
$\beta \sim 0.8$ (e.g.\ Puls et al.\ \cite{puls96}), we do not expect drastic changes of 
the absolute scale from this effect.

An increase of the microturbulent velocity $v_{\rm turb}$ from 5 to 20 km/s
in blanketed models increases the strength of \hei\ (cf. \ Smith \& Howarth \cite{sh98},
\ Villamariz \& Herrero \cite{vh00}), 
and leads to a shift of $\sim$ + 0.05 to 0.1 dex in  $\log(W^\prime)$ 
(i.e.\ towards later types) for models with \teff\ $\la$ 42000 K. 
For hotter stars the difference is negligible.

The effect of line blanketing is strengthened further in denser winds
(cf.\ Abbott \& Hummer \cite{ah85}, Schmutz \cite{s98}).
Models C2 and D2 with an increased mass loss rate by a factor of 2 
show a shift of $\log(W^\prime)$ between $\sim$ -- 0.05 and -- 0.1 dex.

Test calculations for model C2 including also Nickel (Ni~{\sc iv-vi}) show unchanged He lines.
Other models including also Ar, Ne, and Ca confirm that Fe blanketing dominates.

While microturbulence and mass loss affect (though in opposite ways) the 
exact \teff\--scale, their exact importance will have to be studied in future
comparisons.

\subsection{Comparison with observations}
\label{s_compare}

\begin{figure}
\centerline{\psfig{file=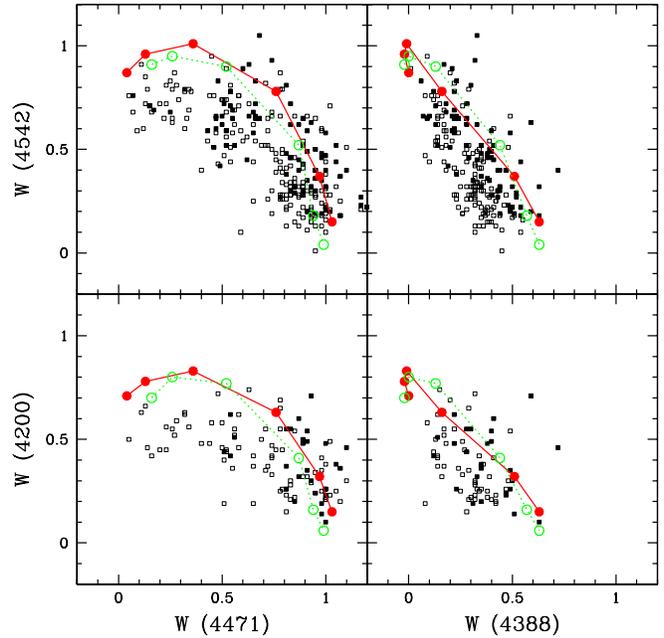,width=9.2cm}}
\caption{Comparison between observed (filled squares: luminosity class V; open squares: 
other luminosity classes) and calculated equivalent widths of 
\hei, \Hei\ $\lambda$4388, \heii, and \Heii\ $\lambda$4200 (in \AA). 
Line blanketed models are indicated by full circles, pure H-He models by open circles.
See discussion in text.
}
\label{fig_obs}
\end{figure}

As a first comparison of our models with observations we show in Fig.\ \ref{fig_obs} 
the predicted and observed equivalent widths of \Hei\ and \Heii\ classification lines
and other strong He lines frequently used in spectral analysis.
The observational data is taken from Mathys (\cite{mat88}, \cite{mat89})
and Conti \& Alschuler (\cite{ca71}).
The observational scatter is real, as the typical measurement errors are $\sim$ 5--7 \%.
The general trend is that the \hei\ and \Hei\ $\lambda$4388 equivalent widths 
are well represented by the models, 
while \heii\ seems to be overestimated by $\sim$ 20\% for spectral types earlier than O7. 
\Heii\ $\lambda$4200 behaves as \heii. 
The other equivalent widths remain essentially unchanged by all other parameter 
variations discussed above (Sect.\ \ref{s_depend}).
A value of $\beta \ga 1.5$, a stronger increase of $\dot{M}$, or an unrealisticly 
large reduction of $\logg$ would be necessary to reduce the predicted equivalent widths 
of the Stark broadened \Heii\ lines.

Strictly speaking, if we were simply to reduce $W(4542)$ by $\sim$ 20\% while keeping 
$W(4471)$ constant for early spectral types, this would result in a change of $\sim$ -- 0.08 
dex in $\log(W^\prime)$ thus reducing the shift in the \teff--scale between line blanketed  
and pure H-He models from $\sim$ 4000 K to 3000 K in the high temperature part.
Future tailored spectral analysis should allow to assess more precisely the
achievable fit accuracy and the precise importance of the parameters discussed in
Sect.\ \ref{s_depend} on the stellar parameters.

\subsection{Comparison with previous analysis}
\label{s_previous}
As discussed in Sect.\ 1, few earlier studies have addressed the effect of line blanketing
in O stars. Essentially all investigations concur with a reduction of \teff\ when
blanketing is included.

Abbott \& Hummer (\cite{ah85}) have constructed a core-halo model where backscattered radiation
due to multiple line scattering in the wind modifies the plane parallel photosphere. 
Their so-called ``wind blanketed'' models yield a decrease of \teff\ by $\sim$ 10 \% for
O4 types (similar to our results), $\sim$ -- 2000 K for an O9.5 supergiant, 
but essentially no shift for O9.5 dwarfs (Bohannan et al.\ \cite{bo90}, Voels et al.\ 
\cite{vo89}). The latter finding is likely due to lack of photospheric blanketing
(inherent to their method) and  modest wind blanketing due to the comparatively 
low mass loss rates of O9.5 dwarfs. 

An improved Monte-Carlo opacity sampling method of a unified photosphere--wind model 
was used by Schaerer \& Schmutz (\cite{ss94}),  
Schaerer \& de Koter (\cite{sdk}), and subsequently applied to a larger parameter 
space by Schmutz (\cite{s98}). 
For mass loss rates comparable to the values adopted here (typical for dwarfs with
low mass loss) the models of Schmutz (\cite{s98}) indicate differences from $\sim$ 
-- 600 K at O8 to $\sim$ -- 2000 K at O4, which is half the shift deduced 
from Fig.\ \ref{fig_scale} and roughly the difference obtained with $Z=1/8$ Z$_{\odot}$ 
(see Sect.\ \ref{s_conclude}). This indicates that their method
underestimates line blanketing compared to {\em CMFGEN}.

Using plane parallel line blanketed non-LTE models based on opacity distribution functions
Hubeny, Heap \& Lanz (\cite{hub98}) found that a pure H-He model with \teff\ $\sim$ 37500 K 
and \logg=4.0 is necessary to reproduce the H and He lines of a line blanketed model with 
\teff\ = 35000 K and same gravity. As can be seen from Fig.\ \ref{fig_scale} our results
are in excellent agreement with their result.

LTE line blocking has been included in plane parallel models by Herrero et al.\ (\cite{her00})
primarily to resolve discrepancies between \Hei\ singlet and triplet lines.
For stars with \teff\ $\ga$ 40000 K this leads to a strengthening of \hei,
opposite to the effect found in all above studies including ours.
This results must be due to an incomplete treatment of the various effects
of line blanketing (cf.\ above), and appears to be unphysical. 
This discrepancy with line blanketed models has also been noted by the authors.

\section{Implications and concluding remarks} 
\label{s_conclude}

The importance of line blanketing obviously depends on metallicity
$Z$. Therefore one may wonder at which $Z$ the stellar parameters will again 
correspond to the results obtained with pure H-He (metal-free) atmosphere
models, i.e.\ close to the VGS96 scale.
Test calculations for models A2 and D2 with a metallicity close to
the SMC value (1/8 \zsun) show still a reduction of \teff\ compared to pure H-He models :
$\Delta$ \teff\ is $\sim$ 60 \% that found at solar metallicity.

As the bolometric correction is essentially unchanged by line blanketing, 
and the $M_V$ versus spectral type ($Sp$) calibration independent to first order
from modeling, we can use the BC-\teff\ relation of VGS96 to derive luminosities
through $\log (L/\lsun) = 2.736 \log\ \teff(Sp) - 0.4 M_V(Sp) - 9.164$.
This relation shows that the predicted reduction of \teff\ by $\la$ 0.04 dex
implies a downward revision of $L$ by $\la$ 0.1 dex for dwarfs of a given spectral type.

Since line blanketing is mostly efficient in the EUV, the ionising spectrum below
912 \AA\ is modified. 
The total number of Lyman continuum photons $Q_0$ predicted by our models
is in good agreement with the calculations of Schaerer 
\& de Koter (\cite{sdk}).
The change of $Q_0$ due to the shift in the \teff-Sp calibration, taking into
account the change of both the radius and the ionising flux per unit surface area $q_0$, 
is given by
$\Delta (\log Q_0) = - 1.264 \Delta (\log \teff) +  \Delta (\log q_0(\teff))$, where
the latter term is dominant (see Schaerer \& de Koter \cite{sdk}).
For a given spectral type between O4V and O9V this amounts typically to a reduction
of $Q_0$ by $\sim$ 40 \%.

While the results presented here provide a clear improvement over earlier
calibrations, and a general reduction of \teff\ due to line blanketing
is unavoidable, we wish to caution that the absolute \teff\ scale
may still be subject to revisions for the following reasons.
First, tailored multi-wavelength analysis of individual objects
are required to test the present models in more depth for O stars,
as recently started by Bouret et al.\ (\cite{bouret_smc}), Hillier et al.
(\cite{john}), and Crowther et al.\ (\cite{cro01}).
Second, the effect of X-rays on the overall ionisation balance and in particular
on the Helium lines remains to be studied. Indeed for late O and B 
stars, depending on the relative X-ray to photospheric flux at
energies close to the relevant ionisation potentials and the wind density,
X-ray emission (likely due to shocks) is expected to increase the
ionisation of most ions (MacFarlane et al.\ \cite{mac94}). Nonetheless, first 
test calculations with {\em CMFGEN} seem to indicate that photospheric lines 
are not affected by X-rays generated in the wind.
Finally, we note that comparisons of photoionisation models
calculated using fluxes from recent atmosphere models (including {\em CMFGEN} 
and Pauldrach et al.\ \cite{p01} models) with {\em ISO} observations
of H~{\sc ii} regions possibly reveal a flux deficiency at energies $\ga$ 
34.8--40.9 eV (Morisset et al.\ \cite{mor01}, but cf.\ Giveon et al.\cite{giv01}).
The importance of the latter two findings --- possibly related to each other ---
on the lines used here as \teff\ indicators remains to be studied.

As UV and optical classification lines of O stars depend in fact 
on several parameters (\teff, gravity, mass loss rate, metallicity, rotation;
e.g.\ Abbott \& Hummer \cite{ah85}, Schmutz \cite{s98}, Walborn et al.\
\cite{wal95}), spectral type and luminosity class calibrations must 
ultimately account for this multi dimensionality.
Some of these issues will be addressed in subsequent publications.

\begin{acknowledgements}
We thank the ``Programme National de Physique Stellaire'' (PNPS) for 
support for this project and the CALMIP center in Toulouse for 
generous allocation of computing time.
\end{acknowledgements}

{}

\end{document}